\documentclass[11pt]{JHEP3}
\usepackage{graphicx}
\usepackage{amssymb}

\def \longg{_{L}}
\def \short{_{S}}
\def\ba{\begin{eqnarray}}
\def\ea{\end{eqnarray}}
\def\be{\begin{equation}}
\def\ee{\end{equation}}
\def\ben{\begin{equation} \nonumber}
\def\een{\end{equation}}
\def\baray{\begin{eqnarray*}}
\def\earay{\end{eqnarray*}}

\def\half{{1\over2}}

\def\D{\mathcal{D}}
\def\({\left(}
\def\){\right)}

\title{Stochastic tunneling for strongly non-Gaussian inflationary theories}
\author{Andrew J. Tolley${}^{\rm a}$ and Mark Wyman${}^{\rm b}$\\
	 Perimeter Institute for Theoretical Physics \\
	 31 Caroline St. N \\
	 Waterloo, ON  N2L 2Y5 \\
	 Canada\\
	 ${}^{\rm a}$ \email{atolley@perimeterinstitute.ca} \, ${}^{\rm b}$ \email{mwyman@perimeterinstitute.ca}}

\abstract{We reconsider the dynamics of stochastic or thermal tunneling in theories like Dirac-Born-Infeld inflation that have non-minimal kinetic terms and, as a result, strongly non-Gaussian perturbations. We first describe a local description of the tunneling process which gives results consistent with the standard Hawking-Moss tunneling. This result is under perturbative control as long as the fluctuation determinant is well approximated by a one-loop integral. We then move to a global description, using the methodology of stochastic inflation and the in-in path integral formalism. This approach shows clearly that the tunneling process becomes strongly coupled whenever the sound speed of the tunneling trajectory departs sufficiently from unity. We argue that these two very different perspectives are nevertheless consistent, and may imply the existence of a simple resummation of the strongly coupled interactions of the field.}

\preprint{PI-COSMO-73}

\begin{document}

\section{Introduction}

Tunneling is the best understood non-perturbative phenomenon in the theory of gravity coupled
to scalar fields.  It is central to
the description of vacuum phase transitions, dynamics on the landscape, and the initial conditions for inflation. There are two types of tunneling in nature: quantum tunneling and thermal tunneling.  Quantum 
tunneling is a classically forbidden process, while thermal tunneling appears for any open system 
whose environment induces a finite temperature and thus can be purely classical.
 In any finite-temperature tunneling system, these processes coexist.
The precise nature of the potential barrier through or over which a system is tunneling 
and the temperature prevalent in the system determine which kind of tunneling is dominant. Thermal tunneling dominates at high temperatures and for very broad potential energy barriers, whereas quantum tunneling dominates for narrow barriers. When considering tunneling in cosmological theories
with a positive potential, the effect of gravity on tunneling is equivalent to adding temperature. This
is because of the finite temperature seen by an observer in de Sitter space.
The presence of gravity thus implies the coexistence of quantum and thermal tunneling in cosmology.

\begin{figure}[htbp] 
   \centering
   \includegraphics[width=4in]{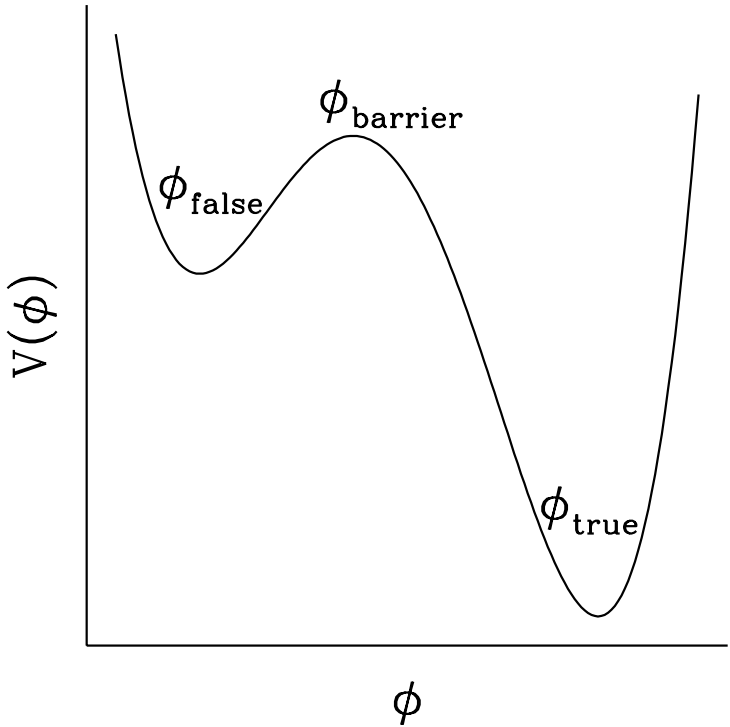} 
   \caption{A typical tunneling potential. The minimum labeled $\phi_{\rm false}$ denotes the false vacuum, while $\phi_{\rm true}$ is the true vacuum, separated by a barrier with maximum at $\phi_{\rm barrier}$.}
   \label{tunnelingpotential}
\end{figure}

Many of the novel inflationary models proposed in the past few years
consist of a scalar field with a non-minimal kinetic term -- at least at the level
of their 4-D effective field theory. The example with the  best physical motivation is Dirac-Born-Infeld (DBI) inflation \cite{DBI}.
In this model, the inflaton is the scalar field describing the location of a D-brane in a warped throat, and the action is of a square root form characteristic of relativistic particle motion. 
More generally, $k$-inflation \cite{kinflation} and $k$-essence \cite{kessence}
are examples of field theories of this type. In such theories, the Lagrangian is an arbitrary function of the scalar field and the usual kinetic term ${\mathcal L}=\sqrt{-g} \, p(X,\phi)$ with $X=-\frac{1}{2} (\partial \phi)^2$.  In fact, all possible predictions of single field inflationary models coming from local theories  can be described within a unified framework using this kind of approach \cite{Creminelli1}. 
In any case, non-minimal kinetic term models
represent explicit, computable alternatives to standard slow roll inflation.

In the case of DBI inflation, quantum tunneling -- i.e., tunneling within the Coleman-de Luccia framework -- has been considered in Ref. \cite{Brown:2007ce}. In this work we shall focus our attention exclusively on thermal tunneling. As we discuss below, there are two ways to describe thermal tunneling: a local description and a global description. Which one we choose depends on whether we wish to study the physics seen by a single observer, or want to ask about the global geometry, including superhorizon modes. The local description is straightforward to generalize to a variety of models, but generalizing the global picture proves to be a more challenging undertaking. There is an immediate, technical challenge: the slow roll approximation cannot be used. This limitation requires a generalization of the stochastic description of inflationary
fields; this we carried out in a previous paper \cite{Tolley:2008na} and was addressed in earlier work \cite{related,related2}. There is, though, a more fundamental hurdle:
the presence of intrinsic non-Gaussianity in these theories plays a crucial role in the tunneling process. 
This may prevent any perturbative calculation from capturing the full phenomenon, forcing us to develop new techniques.

\subsection{Global versus local}

There are two ways to interpret thermal, also known as stochastic, tunneling. To understand these two approaches most clearly, it is helpful to work in the decoupling limit ($M_{\rm pl} \rightarrow \infty$) with fixed Hubble parameter, $H$. In this limit, the scalar fields are taken to exist 
on a fixed de Sitter background geometry; that is, gravity does not respond to the changes
in the scalar fields.
 We begin by considering the local approach.
The local approach describes the tunneling process from the perspective
of an observer within a particular causal patch. 
This patch is covered by what are known as the static coordinates in de Sitter:
\be
ds^2=-(1-H^2r^2)dt^2+(1-H^2r^2)^{-1}dr^2+r^2 d^2 \Omega.
\ee
In this local, or subhorizon, approach, the natural coarse graining of the closed system is:
\begin{itemize}
\item {\bf Open System:} causal patch, sub--horizon modes,  defined in position space as fields with support in $r \le H^{-1}$ 
\item {\bf Environment:} super-horizon modes, 
defined in position space as fields with support in $r > H^{-1}$ 
\item {\bf Tunneling Rate:} Determines rate that the observer's Hubble-sized region ``jumps" over the barrier
\end{itemize}
The only effect of the environment is to induce a finite temperature $T=H/2\pi$ for the subhorizon system~\cite{Gibbons:1977mu}. This idea is known as de Sitter complementarity \cite{Susskind}. Consequently, in this picture tunneling can be described by a finite temperature field theory living in the causal patch.

In the global approach, we consider physics from the perspective of a `superobserver' who can see physics on a global slicing.  For simplicity, we can take this to be the flat slicing in de Sitter,
\be
ds^2=-d\tau^2+e^{2H\tau} d\vec{x}^2.
\ee
In the global, or superhorizon, approach, the natural coarse graining of the closed system is:
\begin{itemize}
\item {\bf Open System:} Long wavelength modes, defined in momentum space as fields with support on
length scales $\lambda > \epsilon^{-1 }aH^{-1}$ \footnote{Here $\epsilon$ is a small parameter introduced so that the the coarse graining scale is a few Hubble radii where the growing mode has already come to dominate over the decaying mode.} 
\item {\bf Environment:} Subhorizon modes, defined in momentum space as fields with support from length scales $\lambda \le \epsilon^{-1 }aH^{-1}$ 
\item {\bf Tunneling Rate:} Determines rate that the field, coarse grained over several Hubble volumes in flat slicing, ``jumps" over barrier 
\end{itemize}
The effect of the environment here is to induce a stochastic noise for the long wavelength system. Adding gravitational dynamics, whilst complicating the description of the system, does not significantly change these considerations.

We stress that these two methods are not calculating precisely the same tunneling rate. Nevertheless, they are sufficiently similar that we expect a comparable result. In brief, they both answer the question: {\it  `What is the rate at which we should expect to find a Hubble volume exiting a metastable vacuum and reaching a stable -- or at least more stable -- vacuum through a stochastic/thermal fluctuation?'}

\section{The local picture: Euclidean Quantum Gravity}

In the local picture, we consider the causal patch to be at finite temperature $T=H/2\pi$. The system is thus described by a partition function
\be
Z={\rm Tr_S} \, e^{-\beta H_{\rm static}},
\ee
with $\beta=2\pi H^{-1}$. The justification for this is as follows. If the global vacuum is chosen to be the Bunch-Davies vacuum $|0\rangle$, then the state that an observer sees is obtained by taking the trace of the density matrix over those modes with support outside the causal patch. This is is found to be equivalent to a thermal state  \cite{Gibbons:1977mu}:
\be
\rho_S = {\rm Tr_L} \( |0 \rangle \langle 0|\) = Z^{-1} e^{-\beta H_{\rm static}}.
\ee
Here $H_{\rm static}(\pi,\phi)$ is the static patch Hamiltonian (written as a function of the field and its spatial derivatives as well as the conjugate momentum) which is the generator associated with the timelike Killing vector in de Sitter. Standard techniques tell us that evaluating this partition function is equivalent to calculating the Euclidean path integral over configurations which are periodic in Euclidean time $\tau_E$ with period $\beta$. That is, we may write
\be
Z=\int D\phi D\pi\, e^{-S_{\beta}(\phi,\pi)}
\ee
where 
\be
S_{\beta}(\phi,\pi)=\int_{-\beta/2}^{\beta/2} d\tau_E\int_0^{H^{-1}} dr \int r^2 d^2\Omega \( \pi \frac{\partial \phi}{\partial \tau_E}- {\cal  H}_{\rm static}(\pi,\phi)\).
\ee
Note that this is a canonical representation of the path integral, which is necessary to get the correct measure in the case of a non-minimally coupled field. For a minimally coupled scalar field, the explicit form of the static Hamiltonian is
\ba
H_{\rm static}&=& \int_0^{H^{-1}} dr \int r^2 d^2  \Omega\, \,  {\cal H}_{\rm static}\\
\nonumber
&=&\int_0^{H^{-1}} dr \int d^2\Omega  \, \( \frac{1}{2}(1-H^2 r^2) \(\frac{\pi^2}{r^2} +r^2\phi_{,r}^2\)+\frac{1}{2}(\nabla_{\Omega}\phi)^2+r^2V(\phi)\) .
\ea
For a configuration in which $\phi$ is fixed at a minimum or a maximum, the energy is 
\be
H_{\rm static} =\frac{4\pi H^{-3}}{3} V(\phi).
\ee
When the field has reached a true equilibrium, the above noted partition function is real: tunneling from the left of a barrier to the right of a barrier is exactly balanced by the reverse process. Nevertheless, we can infer the rate for tunneling from the false to the true vacuum by restricting the range of the phase space path integral so that it only includes configurations with right moving flux and in which the initial state is prepared as being thermal to the left of the barrier and vacuum to the right. This path integral can be evaluated in the saddle-point approximation. Tunneling instantons are those whose fluctuations exhibit an odd number of negative modes. To deal with these, we must perform a Wick rotation of the integration contours. This Wick rotation picks up factors of $i$, and the total tunneling rate is determined by the imaginary part of the partition function
\be
\Gamma \propto {\mathcal Im} \ln Z.
\ee
A class of instantons that always exist are defined by the static configurations
\be
\pi=0, \quad \frac{\delta H_{{\rm static}}(\phi,\pi=0)}{\delta \phi}=0,
\ee
where the field is fixed at a maximum or minimum of its potential. Unlike in Minkowski spacetime, these are guaranteed to have finite action in de Sitter spacetime because the static patch has finite volume.
When the instanton is at a maximum, there is at least one negative mode, hence there is a nonzero contribution to the tunneling rate. In the decoupling limit, the tunneling rate is given by the Arrhenius law
\be
\Gamma \propto \exp{\(-\beta \( H_{\rm static}(\phi_{\rm barrier},0)-H_{\rm static}(\phi_{\rm false},0)\)\)}
\ee
where $\phi_{\rm false}$ is the false vacuum and $\phi_{\rm barrier}$ is the maximum of the potential intermediate between the false and true vacuua. This is precisely the decoupling limit of the Hawking-Moss result
\be
\Gamma \propto \exp{\(S_{\rm ent}(\phi_{\rm barrier})-S_{\rm ent}(\phi_{\rm false})\)},
\ee
where $S_{\rm ent}$ is the entropy of the de Sitter geometry associated with the extrema:
\be
S_{\rm ent}(\phi)=\frac{8\pi^2M_{\rm pl}^2}{H^2(\phi)}=\frac{24\pi^2 M_{\rm pl}^4}{V(\phi)}.
\ee
This follows since
\be
\label{entropychange}
\Delta S_{\rm ent} \sim -\frac{24\pi^2M_{\rm pl}^4}{V^2} \Delta V \sim -2\pi H^{-1} \(\frac{4\pi H^{-3}}{3} \Delta V \)=-\beta \Delta H_{\rm static}.
\ee
In a similar way the Coleman-de Luccia instanton can also be interpreted as finite temperature quantum tunneling~\cite{Brown:2007sd} in the decoupling limit.

\subsection{Extending to non-minimal models}

How does the local picture change in a theory with a non-canonical
relationship between momentum and kinetic energy in its action? Let us consider theories of the general
form, 
\be
S_\phi = \int d^4x \sqrt{-g} \, p(X, \phi). \quad \quad X \equiv - \half (\partial \phi)^2
\ee
The best known example of such a model is DBI inflation. There, the pressure $p$ takes the explicit form
\be
p(X,\phi)=-T(\phi)\sqrt{1-2X/T(\phi)}+T(\phi)-V(\phi),
\ee
where $T(\phi)$ is the tension of the D-brane moving in the warped throat.
The chief novel physics encoded in this action are that 
perturbations travel with a sound speed less than the speed of light.
This sound speed is given by
\be
c_s^2 = \frac{dp}{d\rho} = \frac{p,_X}{p,_X + 2 X p,_{XX}}
\ee
(note that here and throughout the paper we will use commas to indicate partial differentiation;
that is, $P,_X \equiv \partial P /\partial X$).
This means that when we perturb around a cosmological solution, the perturbations travel at a speed typically lower than unity, and there is a sound horizon which is smaller than the causal horizon.
In the local picture we are discussing, this reduced sound speed implies
that Lorentz invariance is broken for the scalar field vacua described by such theories
all the way up to the cut-off energy scale when this effective Lagrangian is expected to breakdown.
To see this, let us consider the short distance singularity structure of the Feynman propagator for fluctuations $\delta \phi$ around some cosmological solution, taken to be in their adiabatic vacuum. Since this will be dominated by the contribution from subhorizon modes, it is sufficient to treat $c_s$ as a constant, since it varies `slowly'. Nevertheless, even high momenta modes travel at the sound speed $c_s$, and so the Feynman propagator will have the Lorentz-violating form
\be
\lim_{{\bf x} \rightarrow {\bf x}'} \langle T\delta \phi({\bf x})\delta \phi({\bf x'}) \rangle = \(\frac{-1}{8\pi^2c_s p_{,X}}\)\frac{1}{(\tau-\tau')^2 - c_s^{-2}\exp(2H\tau)(\vec{x} - \vec{x}')^2 - i \epsilon},
\ee
as if the scalar field perturbations were living on the emergent geometry \cite{emergent}  
\be
ds^2 = (c_s p_{,X})\(-d\tau^2 + c_s^{-2}e^{2H\tau} d\vec{x}^2\).
\ee 
In the case where $c_s$ and $p_{,X}$ are exactly constant, the emergent geometry is also de Sitter, with Hubble constant $H/\sqrt{c_s p_{,X}}$ and associated emergent background temperature equal to $T_{e} = H/(2\pi \sqrt{c_s p_{,X}})$. In the case of a DBI-like kinetic term $p_{,X}=c_s^{-1}$ and so there is no change in the apparent temperature within the emergent geometry. The horizon of the emergent geometry is the sound horizon which the perturbations see and is in general no longer the same as the true event horizon.

Nevertheless, the physical temperature of the background is still the same. Following the argument of Ref.~\cite{Gibbons:1977mu}, if the Feynman propagator is defined to be analytic in the emergent geometry for spacelike separated points then it will be invariant under
\be
\tau \rightarrow  \tau + i \frac{2\pi}{H},
\ee
which is the same as requiring analyticity of the propagator in the real geometry.
This is the key step in justifying the notion that subhorizon physics in de Sitter is equivalent to a finite temperature system of temperature $T=H/2\pi$. This is true when $c_s$ and $p_{,X}$ are constant because the emergent metric depends on time only through $e^{2H\tau}$, which is invariant under this shift.

What this example illustrates is that even if we choose the global state to be the adiabatic vacuum state for fluctuations around nontrivial cosmological solution, with nontrivial speed of sound, i.e. even if we choose the global vacuum {\it not} to be Bunch-Davies, it still makes sense to think of the subhorizon system as being at a finite temperature $T=H/2\pi$ set by the event horizon and {\it not} the sound horizon.

In short, causality demands that the part of the global vacuum describing superhorizon physics is never relevant to a local observer. 
Let us denote  the Bunch-Davies vacuum by $|0\rangle$.  At a given time, the
apparent vacuum $|\alpha\rangle$ for perturbations around some cosmological solution formally corresponds to a coherent state\footnote{Technically speaking, the coherent state may be in a unitarily inequivalent representation since the sum $\sum_i |\alpha_i|^2$ may not converge in the UV.  In the context of field theory with a cutoff, this distinction is irrelevant, and is really a reflection of the fact that we are not allowed to excite modes lying beyond the cutoff. We take the point of view that these manipulations are meaningful if, having defined them for fixed cutoff, they make sense in the limit that the cutoff tends to infinity, after appropriate renormalizations.}
\be
| \alpha \rangle = U | 0 \rangle = e^{-|\alpha|^2}\exp [\alpha a  + \alpha^\dag a^\dag ] | 0 \rangle,
\ee
where the $\alpha$ and $\alpha^\dag$ are defined for the particular state, and $\alpha a$ is shorthand for a sum over a complete set of modes $\sum_i \alpha_i a_i$. 
We can always divide the creation operators into superhorizon (L) and subhorizon (S) modes,
\be
a = a_L + a_S,
\ee
such that we guarantee $[a_L,a_S]=[a_L,a^{\dag}_S] = 0$. As a density matrix this state is
\be
|\alpha \rangle \langle \alpha | = U  | 0 \rangle \langle 0 | U^{-1} = U_L U_S  | 0 \rangle \langle 0 | U_S^{-1} U_L^{-1}. 
\ee
where $U_S=e^{-|\alpha_S|^2}\exp [\alpha_S a_S  + \alpha_S^\dag a_S^\dag ]$ and similarly for $U_L$.
A local observer can only see the part of the state obtained by tracing out the long wavelength modes, namely
\begin{eqnarray}
\rho_S={\rm Tr_L} | \alpha \rangle \langle \alpha | & = & {\rm Tr_L} (  U_L U_S  | 0 \rangle \langle 0 | U_S^{-1} U_L^{-1} )\\
& = &  U_S ({\rm Tr_L} U_L | 0 \rangle \langle 0 | U_L^{-1}) U_S^{-1}= U_S ({\rm Tr_L} | 0 \rangle \langle 0 | ) U_S^{-1}\\
& = &U_S   Z^{-1} e^{-\beta H_{\rm static}}  U_S^{-1}=Z^{-1} e^{-\beta U_S H_{\rm static}U_S^{-1}}.
\end{eqnarray}
In other words, if the global vacuum is the adiabatic vacuum defined around some cosmological solution, the state seen by a local observer is a coherent excitation of a thermal state with the same temperature $T=H/2\pi$. 

Nothing about these changes implies that the usual thermal tunneling
instanton should no longer exist or be substantially changed. By construction,
the thermal instanton in this picture occurs via a static field configuration. 
Since the novel effects from the nonminimal kinetic term only arise when velocity is finite
(for instance, in DBI inflation, $p(X,\phi) \simeq X - 2X^2/T(\phi)-V(\phi)$ for $\dot{\phi} \ll \sqrt{T}$), 
there is little influence of the kinetic term on the static configuration.
Thus, the instanton itself
and the fluctuations around it -- which are necessary to determine the existence of a negative mode and hence the existence of the instanton -- are not affected by the nonminimal kinetic terms. 

Any difference between the nonminimal and minimal cases will come from higher-order corrections, specifically two-loop corrections in the Euclidean path integral. The instanton calculation is a saddle-point calculation, usually considered to one-loop order, that is only valid as long as the two-loop and higher corrections are negligible. Fluctuations around the instanton generate terms in the Lagrangian schematically of the form
\be
\delta {\mathcal L} = p_{,X} \delta X +\frac{1}{2}p_{,XX}\delta X^2 +\frac{1}{3!}p_{,\phi\phi\phi} \delta \phi^3+p_{,X\phi}\delta X \delta \phi =\dots
\ee
where $\delta X=-\frac{1}{2}(\partial \delta \phi)^2$.
A necessary, but not sufficient, condition for the one-loop calculation to be valid is
\be
\delta X \ll \left. \frac{p_{,X}}{p_{,XX}} \right|_{\phi=\phi_{\rm barrier}, \pi=0},
\ee
since at the point of equality the fluctuations in the higher order contributions to the kinetic term become as large as the quadratic part. 
For a minimally coupled scalar field $p_{,XX}=0$ and so this condition is always satisfied. 
In the language of instantons, when this condition is not satisfied, the fluctuation determinant can no longer be determined by a gaussian integral about the instanton solution.

The typical magnitude of the fluctuations is set by the Hubble scale $\delta \phi \sim H$, $\delta \dot{\phi} \sim H^2$. 
In the case of a DBI model, this condition for the breakdown of the one-loop approximation
is, then,
\be
\frac{ \delta X}{T(\phi_{\rm barrier})} \sim \frac{\delta \dot{\phi}^2}{T(\phi_{\rm barrier})} \sim \frac{H^4}{T(\phi_{\rm barrier})} \sim 1.
\ee
This result is intuitive. 
We expect loop corrections to the effective action to be of order at least $H^4$. 
On the other hand $T(\phi)$ sets the scale of the kinetic term for DBI, so this is just the condition for the loop corrections to the effective action to become comparable to the classical kinetic term. This is, of course, not the only condition for the validity of the one-loop approximation. However, the additional conditions have to do with the $\phi$ dependence $p(X,\phi)$ and so are typically easier to satisfy if the potential is sufficiently broad.

This is similar but not the same as the perturbative bounds that have been placed on eternal inflation, when quantum fluctuations dominate over classical evolution in the field evolution in Refs.~\cite{Leblond:2008gg, Creminelli:2008es}. The key difference is that in our case we are expanding around a static solution (the instanton), whereas in these other works the authors have found nontrivial bounds by expanding around a time dependent solution with $c_s \neq 1$. The two calculations are qualitatively similar in that they reflect the break down of perturbation theory for the fluctuations -- in one case around the instanton, in the other the cosmologically rolling solution -- but should not be expected to match in detail.

We have thus established a subset of cases where the thermal tunneling instanton is not valid. It will be a bad approximation when the top of the potential
barrier is at such a high energy that all perturbations are dominated by relativistic effects. In contrast, the Hawking-Moss instanton for a minimally coupled scalar field is valid  \cite{Batra:2006rz} even when the top of the potential barrier reaches the energies
associated with eternal inflation.  This is closely connected with the observation that a minimally coupled field
reaches eternal inflation within the regime of perturbative control \cite{Leblond:2008gg}, but a non-minimal theory such as DBI inflation is not under perturbative control in the regime of eternal inflation, at least when the sound speed departs sufficiently from unity. The difference is that the loop expansion for a minimally coupled field is determined by the potential in the decoupling limit, rather than
by its kinetic term. If the potential is sufficiently flat, then these corrections are always negligible.
This is always true when slow roll is a good approximation, and can be true even at the top of a broad potential barrier in a tunneling set up.
 
We now turn to the complementary, global picture
of de Sitter space and inflation, where the techniques of stochastic inflation
can help us to understand better what exactly is going on in thermal tunneling.

\section{The global picture: stochastic inflation}

In any accelerating geometry, perturbations of a quantum field expand faster than the Hubble radius. 
As these perturbative modes cross the Hubble horizon, 
their two quantum oscillating components split into a growing and a decaying classical mode. 
The long wavelength state of the quantum field  retains any non-linearities it possesses, but is squeezed by the expansion and becomes
progressively more classical as its growing mode dominates over its decaying mode. 
The long wavelength or background field evolution thus has two contributions: part from its 
overall classical evolution, part from the stochastic kicks it receives as each subhorizon mode 
exits the horizon and adds itself to the collection of superhorizon modes constituting the 
background.

This process is always operative for inflating geometries, but
the stochastic influence of the subhorizon noise is generally small enough to ignore. However,
the process of thermal tunneling is precisely the exception that proves this rule: a stochastic
jump of rare vigor comes along that is of sufficient size to move the average field at its false vacuum value
in one Hubble patch over the potential barrier
that separates it from the true vacuum.
 The other time when large perturbations are important is, of course, during eternal inflation. 
 Both of these phenomena can be described within the framework of stochastic inflation.

\subsection{Deriving the stochastic formalism}

A rigorous derivation of the stochastic approach is quite involved, but we can sketch the details as follows: Consider the evolution of the system in the flat slicing with canonical action
\be
S[P,\phi]=\int d\tau d^3 x \, e^{3H\tau} \(P \frac{\partial{\phi}}{\partial{\tau}}-{\cal H}_{\rm flat}(P,\phi) \),
\ee
The momentum conjugate $P$ defined here should not be confused with $\pi$ which was the momentum in the static slicing.
For instance, for the case of a massive minimally coupled scalar field on de Sitter we have
\be
{\cal H}_{\rm flat}(P,\phi) = \frac{1}{2}P^2 + \frac{1}{2a^2}(\nabla \phi)^2 + \frac{1}{2}m^2 \phi^2,
\ee
with $a(\tau)=e^{H\tau}$.
Expectation values of fields can be derived from considering the {\it in-in} or closed-time-path integral, which contains two branches $+$ and $-$. For instance we have
\be
\label{inin1}
\langle 0 | T e^{i\int J \phi}| 0\rangle= \int D\phi_+ DP_+D\phi_- DP_- e^{iS[P_+,\phi_+]-iS[P_-,\phi_-]+i\int J\phi_+},
\ee
and the path integral is performed with final time boundary conditions $\phi_+(\tau_f)=\phi_-(\tau_f)$, $P_+(\tau_f)=P_-(\tau_f)$.

To connect with the usual stochastic formulation, we want to construct a probability distribution on phase space. In an appropriate limit, this approach will give rise to a Fokker-Planck equation describing the stochastic evolution. In quantum theory,  for a density matrix $\rho$ we can define a quasi-probability distribution -- the Wigner function -- as follows
\be
W[\phi,P; \tau]=\int D \mathcal{E} \, e^{i\int d^3 x e^{3H\tau }P\mathcal{E} } \langle \phi-\frac{\mathcal{E}}{2}| \rho(\tau)|\phi+\frac{\mathcal{E}}{2} \rangle.
\ee
As is well known, this function can be used to calculate expectation values in the normal way as if it were a probability distribution defined on phase space. The quantum information is contained in the fact that $W$ is not positive definite.

The Wigner function can also be thought of as the `wavefunction' associated with the functional Fourier transform of the path integral (\ref{inin1}), given the replacements $\phi_{\pm}=\phi\mp\mathcal{E}/2$ and $P_{\pm}=P \pm \chi/2$
\be
W[\phi,P]=\int^{\phi} \D\phi \int^{P} \D P \int \D\mathcal{E} \int \D \chi \exp \({i \hat{S}}\),
\ee
with 
\ba
\hat{S} &= \int d\tau d^3x e^{3H\tau} & \left[ \chi \partial_{\tau}\phi +\mathcal{E} \( \partial_{\tau}P+3HP\)-{\cal H}_{\rm flat}[\phi-\frac{\mathcal{E}}{2},P+\frac{\chi}{2}] \right. \nonumber \\
&& \left. \quad +{\cal H}_{\rm flat}[\phi+\frac{\mathcal{E}}{2},P-\frac{\chi}{2}] \right].
\ea
Just as in the local approach, we traced out the superhorizon degrees of freedom to determine the subhorizon density matrix, here we define the long wavelength density matrix $\rho \longg$ by separating each field $A=(\phi,P,\mathcal{E},\chi)$ into long and short wavelength modes
\ba
&&A(x)=A \longg(x)+A \short(x) ,\\
&&A_L =\int \frac{d^3k}{(2\pi)^3} e^{ik.x}\phi(k,\tau) W(k;\tau) , \quad A_S =\int \frac{d^3k}{(2\pi)^3} e^{ik.x}\phi(k,\tau) (1-W(k;\tau)).
\ea
Here $W(k;\tau)$ is the Window function that distinguishes super and sub-Hubble modes, and is most commonly taken to be of the form
\be
W(k;\tau)=\theta(\epsilon a H -|k|),
\ee
with $\epsilon$ a constant whose value is taken to be much less than unity. This is in fact a crucial point: the coarse graining is not performed at the Hubble horizon, because the state still appears quantum there. The state only appears classical once the squeezing mechanism has taken effect, when the modes are a few e-folds outside the Hubble horizon.
This split is defined so that equal time bilinears of the form (for some functions $A$ and $B$) $\int d^3x A\longg(x,\tau) B\short(x,\tau) =0.$
However since the coarse graining is explicitly time dependent, we have
\ba
\int d^3x A\longg \partial_{\tau} B\short &=&-\int d^3x \( \partial_{\tau}A\longg\)  B\short=-\int \frac{d^3k}{(2\pi)^3} A(-k)B(k)W(k;\tau) \partial_{\tau}W(k;\tau)  \nonumber \\
&=&- \frac{\epsilon aH^2 }{2}\int \frac{d^3k}{(2\pi)^3} A(-k)B(k) \delta(|k|-aH).
\ea
This means that independent of the form of the Hamiltonian, there is always a direct coupling between the momentum and field of the short and long wavelength modes through the $P \partial_{\tau} \phi$ term in the Lagrangian. For a free field theory, this is the only coupling term.

We construct the reduced Wigner function describing the effective long wavelength dynamics by tracing out the short wavelength modes
\be
W_r\left[\phi\longg,P\longg\right]=\int \D\phi\short \ \D P\short \ W\left[\phi\longg+\phi\short,P\longg+P\short\right]\,.
\ee
The reduced Wigner function is the `wavefunction' associated with the path integral,
\be
W_r\left[\phi\longg,P\longg\right]=\int \D\phi\longg \ \D P\longg \  \D\mathcal{E}\longg \ \D\chi\longg\, 
\exp\({i  \hat S \left[\phi\longg,P\longg,\mathcal{E}\longg,\chi\longg\right]}\) \, F\left[\phi\longg,P\longg,\mathcal{E}\longg,\chi\longg\right] \label{rwig}
\ee
where $F$ is the Feynman-Vernon infuence functional
\be
F\left[\phi\longg,P\longg,\mathcal{E}\longg,\chi\longg\right]= \int \D\phi\short \ \D P\short \  \D\mathcal{E}\short \ \D\chi\short
\exp\({i  \hat S \left[\phi,P,\mathcal{E},\chi\right]-i \hat S \left[\phi\longg,P\longg,\mathcal{E}\longg,\chi\longg\right]}\) \,.
\ee
The full expression for $F$ is obviously difficult to determine, however in most cases of physical interest, $\log F$ has a controlled perturbative expansion in powers of $\mathcal{E}\longg$ and $\chi\longg$. We will not require this sort of expansion for our calculations. However, let us 
write down concretely what form such an expansion would take. Since we have 
$F^*\left[\phi\longg,P\longg,\mathcal{E}\longg,\chi\longg\right]=F\left[\phi\longg,P\longg,\mathcal{E}\longg,-\chi\longg\right]$, we can simply expand $\log F$ in the following way:
\ba
\label{influenceexpansion}
\hspace{-10pt}\log F&=&C\left[\phi\longg,P\longg\right]+\int \( i A_\phi \left[\phi\longg,P\longg\right]\mathcal{E}\longg
+i A_P \left[\phi\longg,P\longg\right]\chi\short \) e^{3H\tau}d^3x d \tau\\
&&\hspace{-5pt}-\int \hspace{-5pt}\int d^4x  d^4ye^{3H (\tau_x+\tau_y)}\(\frac 12 \mathcal E\longg (x)D_{\phi\phi}(x,y) \mathcal{E}\longg(y)
+\frac 12 \mathcal{E}\longg(x)D_{\phi P}(x,y) \chi\longg(y)+ \nonumber \right.  \\ 
&&\left. \frac{1}{2} \chi\longg(x)D_{PP}(x,y) \chi\longg(y)
  \right ) \nonumber
+\mathcal{O}(\mathcal{E} ^3, \chi^3, \mathcal{E}^2\chi, \mathcal{E}\chi^2)\nonumber\,,
\ea
For some functions $C$, $A_X$, and $D_{XX}$ that will encode the particularities of the
system one is studying (where $X$ stands for the field or momentum, as above) . The influence functional can also be represented by its functional Fourier transform 
\ba
F\left[\phi\longg,P\longg,\mathcal{E}\longg,\chi\longg\right]=\int \D \eta_\phi\D\eta_P {\cal P}\left[\eta_\phi,\eta_P,\phi\longg,P\longg\right]
e^{-i \int\left[\eta_\phi \mathcal{E}\longg+\eta_p\chi\longg\right]e^{3H \tau} d^3x d \tau}\,,
\ea
with $\eta_\phi$ and $\eta_P$ being the conjugate variables to $\mathcal{E}_L$ and $\chi_L$.
We will see soon that these $\eta_X$ will function as noise forces in the equations
of motion for our system.
The advantage of this approach can now be seen. In the semi-classical limit we can approximate
\ba
\label{longclassical}
\hat S\left[\phi\longg,P\longg,\mathcal{E}\longg,\chi\longg\right]\approx 
\int d^3xd\tau e^{3H \tau}\left[\chi\longg\partial_\tau \phi\longg+ \mathcal{E}\longg\left[\partial_\tau +3 H\right] P \longg
+\mathcal{E}\longg \frac{\delta H\longg}{\delta \phi}-\chi\longg\frac{\delta H\longg}{\delta P}
\right]\,,\nonumber
\ea
where $H_L=\int d^3 x \, {\cal H}_{\rm flat}(\phi_L,P_L)$.
We can then easily perform the path integral (Eqn. \ref{rwig}) over $\mathcal{E}\longg$ and $\chi\longg$ with the result
\ba
W_r\left[\phi\longg,P\longg\right]&\approx&\int^{\phi \longg}\D \phi\longg\, \int^{P \longg} \D P\longg\, \int \D \eta_\phi \int \D \eta_P 
\, \, {\cal P}\left[\eta_\phi,\eta_P,\phi\longg,P\longg\right] \nonumber \\
&\times &\delta\left[\partial_\tau \phi\longg-\frac{\delta H\longg}{\delta  P}-\eta_\phi\right]
 \delta\left[\partial_\tau P\longg+3 H P\longg+\frac{\delta H\longg}{\delta \phi}-\eta_P\right]\,. \label{finalwig}
\ea
Thus we see that the reduced Wigner function is the classical probability distribution associated with the two-noise Langevin equations, as considered in reference~\cite{Tolley:2008na}
\ba
\partial_\tau \phi\longg&=&\frac{\delta H\longg}{\delta P}+\eta_\phi\\
\partial_\tau P\longg&=&-3 H P\longg-\frac{\delta H\longg}{\delta \phi}+\eta_P\,,
\ea
where ${\cal P}$ in Eqn. \ref{finalwig} is a quasi-probability distribution that encodes the ``coloring"  -- the statistical behavior -- of the noise terms.

\subsection{Irreversibility and the fluctuation theorem}

Having connected with the approach of~\cite{Tolley:2008na} we can now address the question of how to calculate the tunnelling rate. The key observation is that the dominant contribution to the tunneling rate is determined by the irreversible nature of the coarse-grained dynamics.
A crucial difference between the stochastic approach and the local approach of the previous section is that in the stochastic case the effective dynamics of the long wavelength system is irreversible due to the presence of dissipation (Hubble damping) and diffusion (noise from subhorizon quantum fluctuations).  In the local case causality guarantees that the superhorizon and subhorizon modes decouple, and hence there is no diffusion, and since the static time coordinate is a Killing vector, there is also no damping seen by a static observer. 

To be more precise about what we mean by irreversibility, consider the propagator ${\cal K}$ that evolves the reduced Wigner function between two times $\tau_i$ and $\tau_f$
\be
W_r\left[\phi^f\longg,P\longg^f,\tau_f \right]=\int \D\phi^i\longg \D P^i\longg \, \, {\cal K}\left[ \phi^f\longg,P\longg^f,\tau_f;\phi^i\longg,P\longg^i,\tau_i\right]W_r\left[\phi^i\longg,P\longg^i,\tau_i \right].
\ee
The propagator ${\cal K}$ has the path integral representation
\ba
&&{\cal K}\left[ \phi^f\longg,P\longg^f,\tau_f;\phi^i\longg,P\longg^i,\tau_i\right] = \\
\nonumber
 &&\int_{\phi\longg(\tau_i)=\phi^i\longg}^{\phi\longg(\tau_f)=\phi^f\longg}  \D\phi \int_{P\longg(\tau_i)=P^i\longg}^{P\longg(\tau_f)=P^f\longg}  \D P  \int \D\mathcal{E}\int \D\chi\longg
\, \exp\({i  \hat S \left[\phi\longg,P\longg,\mathcal{E}\longg,\chi\longg\right]} \) \, F\left[\phi\longg,P\longg,\mathcal{E}\longg,\chi\longg\right]
\ea
The fundamental statement of the irreversibility of the system can be encoded in the following ratio
\be
{\cal R}\left[ \phi^f\longg,P\longg^f,\tau_f;\phi^i\longg,P\longg^i,\tau_i\right]=\frac{{\cal K}\left[ \phi^f\longg,P\longg^f,\tau_f;\phi^i\longg,P\longg^i,\tau_i\right] }{{\cal K}\left[ \phi^i\longg,-P\longg^i,\tau_f;\phi^f\longg,-P\longg^f,\tau_i\right] }
\ee
In other words, for a given transition between two states $A \rightarrow B$, we can compare the probability for this transition, with the probability for the time reversed transition $B \rightarrow A$ to occur in the same time. For a reversible system, we have ${\cal R}=1$, and so $\cal{R}$ encodes any departure from reversibility, which in practice is a measure of the amount of dissipation and diffusion.

A priori, this ratio can be a complicated function of the initial and final phase space values and times. However, in many well understood systems, the ratio takes the simple {\it detailed balance} form
\be
\label{detailedbalance}
{\cal R}=\exp\({M[\phi_L^f,P_L^f]-M[\phi_L^i,P_L^i]}\),
\ee
for some function, $M$. 
This result is closely connected to a result known in the condensed matter literature as the {\it fluctuation theorem}~\cite{FT}. It states that the ratio of the probabilities of making a transition from $A \rightarrow B$ to the ratio for the time reversed process $B \rightarrow A$, in some time $\tau_f-\tau_i$ is the exponential of the amount of entropy produced in the transition:
\be
{\cal R}=\frac{P(A\rightarrow B;\tau_f-\tau_i)}{P(B \rightarrow A;\tau_f-\tau_i)}=\exp\( \int_{\tau_i}^{\tau_f} d\tau \frac{\partial{S_{\rm ent}}}{d\tau}\)=\exp \( S_{\rm ent}(B)-S_{\rm ent}(A)\),
\ee
where $S_{\rm ent}$ is the entropy associated with the system. Thus it is natural to identify $M[\phi_L^i,P_L^i]$ as the entropy associated with a given phase space configuration.

The fluctuation theorem is the modern statement of the Second Law of Thermodynamics, encoding the fact that the probability for entropy to increase is greater than that for it to decrease, recognizing that whilst spontaneous entropy decrease is allowed, its probability is highly suppressed.
This exponential suppression for the reverse process, directly determines the exponential suppression of the stochastic tunneling rate. The rate to go from the false vacuum to the top of the barrier is given by the exponential of the de Sitter entropy difference between the two configurations multiplied by the rate to go from the top of the barrier to the false vacuum:
$P({\rm false \rightarrow barrier})=\exp(S_{\rm ent}(\phi_{\rm top})-S_{\rm ent}(\phi_{\rm false}))P({\rm barrier \rightarrow false})$. The latter process $P(\rm barrier \rightarrow false)$ though, is a classically allowed process; hence its transition probability will contain no additional exponential suppression. Similarly, once the field has reached the top of the barrier it can roll classically down towards the true vacuum. Thus we infer that the tunneling rate is given by
\be
\Gamma({\rm false} \rightarrow {\rm true}) \propto  \exp{\(S_{\rm ent}(\phi_{\rm barrier})-S_{\rm ent}(\phi_{\rm false})\)},
\ee
up to factors determined by the classically allowed processes.

So, how does this connect with cosmology? We have performed the above analysis for inflation with a minimally coupled scalar field. We have expanded  the influence functional only to quadratic order as in equation [\ref{influenceexpansion}]. In this case, one can explicitly show that in the ultralocal limit, the detailed balance condition [\ref{detailedbalance}] holds, and furthermore that it is consistent with the fluctuation theorem:  the change in $M$ is precisely equal to the decoupling limit of the change in the entropy (eq.~[\ref{entropychange}]) of a scalar field in a de Sitter spacetime determined by the potential $V(\phi)$. Even when gravitational backrection is included, this result holds up~\cite{Tolley:2008na,Lindebook}.
In other words, the dominant contribution to the stochastic tunneling rate is identical to the rate determined by Hawking and Moss in the local case via the instanton method. This strongly supports the interpretation of the Hawking-Moss transition as being equivalent to the stochastic tunneling process, as is usually assumed~\cite{Lindebook}.

Our interpretation of this result is that the two results agree in the usual case (Hawking-Moss and stochastic) precisely because of thermodynamic reasons. In the Hawking-Moss case this is explicit, because it is a thermal calculation of the imaginary part of the free energy. In the stochastic case, it is explicit because the evolution of the Wigner function is consistent with detailed balance and the fluctuation theorem. 

\subsection{Irreversibility for DBI inflation, non-gaussian corrections to tunneling rate}

In a previous paper~\cite{Tolley:2008na} we extended the stochastic formalism to DBI inflation, and in the language of this section, derived the Langevin/Fokker-Planck equations that follow expanding the log of the influence functional to quadratic order as described in equation [\ref{influenceexpansion}], and then taking the ultralocal long wavelength limit. We found that to this order, the evolution of the (Wigner) probability distribution on phase space, indeed satisfied detailed balance, in the form
\be
{\cal R}=\exp\({M[\phi_L^f,P_L^f]-M[\phi_L^i,P_L^i]}\),
\ee
but that the full expression for $M$ was 
\be
M[\phi,P]=M[\phi,J]=-16 \pi^2 \int d\phi \, M_{\rm pl}^2 \(\frac{c_s H_{,\phi}(\phi,J)}{H(\phi,J)^3} \) 
\ee
where $J$ is related to $\phi$ and $P$ via the canonical transformation used to derive the Hamilton-Jacobi formalism~\cite{Tolley:2008na}. As pointed out in Refs.~\cite{Tolley:2008na,Wohns:2008qv} if this result is taken at face value, the presence of the additional factor of $c_s$ inside the integrand means that the associated stochastic tunneling rate, whose suppression is by the above arguments determined by ${\cal R}$, is no longer given by the same result as the local thermal tunneling rate, since $M$ is no longer identical to the expression for the entropy of de Sitter spacetime, with the associated Hubble constant $H(\phi,J)$. It is only in the case $c_s=1$, in other words in the minimally coupled scalar case do we recover the usual answer.

The stochastic tunneling rate and the Hawking-Moss tunneling rate for models in which $c_s$ may depart from unity along the tunneling path thus appear to disagree.  One might think that this apparent discrepancy should be taken as evidence that the two rates do not describe quite the same thing. We find that result unlikely since, as we explained at length in the introduction, they both effectively describe the rate for a Hubble sized region to traverse the barrier between the false and true vacuum. If this is not the case, then it must be that the approximations used in deriving one of the two results breaks down. In the local instanton method, we have already argued that the calculation is usually well under control from higher loop corrections to the tunneling rate, as long as the condition $H^4 \ll T(\phi_{\rm barrier})$ is satisfied. This suggests that it is the stochastic calculation that is failing. 

Let is consider the approximations used in deriving the stochastic formalism, sketched out above. The key approximations are:
\begin{itemize}

\item{{\it Long wavelength perturbations are treated as classical}: This is encoded in equation [\ref{longclassical}] where $\hat{S}$ is expanded to linear order in ${\cal E}_L$ and $\chi_L$.}

\item{{\it Long wavelength modes decouple, i.e.~the ultralocal limit}: In this approximation we can neglect spatial gradients for the long wavelength fluctuations, and the noise correlators are assumed to be local in space $\langle \eta(x)\eta(x') \rangle \propto \delta^3(x-x')$. This is explained in more detail in Ref.~\cite{Tolley:2008na}.}

\item{{\it Short wavelength noise can be treated as gaussian}: This is encoded in equation [\ref{influenceexpansion}]} where the influence functional is taken to be gaussian in ${\cal E}_L$ and $\chi_L$.)

\end{itemize}

The first and second properties follow from the squeezing of the long wavelength quantum state and causality respectively. Both of these properties are equally well satisfied in DBI inflation as they are in usual slow roll inflation. However, the third approximation is not, precisely because in DBI inflation, the quantum fluctuations in the inflaton are more non-gaussian, then in turn the stochastic noise which arises from the short wavelength fluctuations is more non-gaussian. 

Our central claim is that it is the neglect of the non-gaussian noise fluctuations that is responsible for the apparent discrepancy between the stochastic tunneling rate and the Hawking-Moss result. As soon as $c_s$ departs noticeably from unity along the tunneling trajectory, the non-gaussian contributions to the irreversibility ratio ${\cal R}$ become important. Indeed it seems plausible that the physical consistency of these two results requires that the ratio ${\cal R}$ is always consistent with the fluctuation theorem, and so when the corrections from the non-gaussian noise fluctuations are resummed, we will recover the usual result.

To restate this problem for DBI inflation, whenever $c_s$ departs from unity along the tunneling trajectory, perturbation theory in the variables ${\cal E}_L$ and $\chi_L$ which are conjugate to the noise in the in-in path integral breaks down. The path integration over these variables must be performed non-perturbatively. Unfortunately it is not clear how to achieve this as yet.

To see that the noise fluctuations are becoming non-perturbative during tunneling, let us compute the three-point function for the noise $\eta_{\phi}$. Following the coarse-graining recipe described above, and in more detail in Ref.~\cite{Tolley:2008na}, the noise $\eta_{\phi}$ is related to the subhorizon fluctuations as follows (note: in the ultralocal limit we can drop the $x$ dependence)
\be
\eta_{\phi}(\tau) = -\int \frac{d^3 k }{(2\pi)^3} a(\tau)H^2 \delta (|k|-\epsilon aH) \phi_S(k).
\ee
where $\phi_S$ are the short wavelength fluctuations about the long wavelength background $\phi_L$.
The key point is that the noise has a non-zero three-point correlation function, precisely because the subhorizon fluctuations have a non-zero three-point function.
As we are only interested in the order of magnitude, we can estimate the three-point function as follows: in the framework of cosmological perturbation theory we would define the comoving curvature perturbation $\zeta \sim H\phi_S/\dot{\phi}_L$. It is usual to encode the non-gaussianity of the the fluctuations in terms of the parameter $f_{NL}$ which is in general scale and shape dependent, 
\be
\langle \zeta^3 \rangle  \sim -\frac{3}{5} f_{NL} \( \langle \zeta \rangle\)^2.
\ee
In terms of the fluctuations $\phi_S$ this is
\be
\langle \phi_S^3 \rangle  \sim -\frac{H}{\dot{\phi}_L} f_{NL} \( \langle \phi_S \rangle\)^2.
\ee
Given the expression for the noise in terms of the fluctuations $\phi_S$, we find the three-point function for the noise to be of the form
\be
\langle \eta_{\phi}(\tau_1) \eta_{\phi}(\tau_2)\eta_{\phi}(\tau_3) \rangle \sim f_{NL}\frac{H^6}{\dot{\phi}_L} \delta(\tau_1-\tau_2)\delta(\tau_1-\tau_3),
\ee
where we have made the simplifying assumption that the mixing of modes that exit the horizon at different times is negligible, which provides the two delta functions. This should be compared with the form for the two-point function for the noise
\be
\langle \eta_{\phi}(\tau_1) \eta_{\phi}(\tau_2)\rangle \sim H^3 \delta(\tau_1-\tau_2).
\ee
Consider the dimensionless noise variable $q(\tau)=\frac{\eta_{\phi}}{\dot{\phi}_L}$
which encodes the ratio of the stochastic fluctuations to the classical evolution. We can average this over the tunneling trajectory that takes time $\Delta \tau$
\be
\bar{q} = \Delta \tau^{-1}\int_0^{\Delta \tau} d\tau q(\tau)
\ee
then a simple order of magnitude estimate gives
\be
\langle \bar{q}^2\rangle \sim \frac{H^3}{\dot{\phi}_L^2} \Delta \tau
\ee
and
\be
\langle \bar{q}^3 \rangle \sim \langle f_{NL}\rangle \frac{H^6}{\dot{\phi}_L^4} \Delta \tau^2
\ee
Here we understand $\langle f_{NL} \rangle$ to be the average of $f_{NL}$ along the tunneling trajectory.
In other words just as $\langle f_{NL} \rangle $ naturally encodes the nonlinearities in $\zeta$, it also naturally encodes the non-linearities in $\bar{q}$
\be
\langle \bar{q}^3 \rangle \sim \langle f_{NL}\rangle \( \langle \bar{q}^2\rangle\)^2.
\ee
Now the key point is that since $\bar{q}$ measures the ratio of the stochastic fluctuations to the classical evolution, during tunneling, which is a rare quantum event where the quantum fluctuations  are sufficient to dominate over the classical evolution, we inevitably have $\bar{q} \sim 1$. But unless $\langle f_{NL}\rangle \ll 1$, for any such large fluctuation all the higher order correlation functions of $\bar{q}$ become of order unity, and hence perturbation theory breaks down in the noise variable. 

We have explicitly checked by a more detailed calculation that the above argument goes through when the corrections to the irreversibility ratio ${\cal R}$ are computed from the non-gaussian noise fluctuations. The bound on the perturbative region we find for tunneling, namely $\langle f_{NL} \rangle \ll 1$, is much more restrictive than that discussed
in Refs.~\cite{Leblond:2008gg, Creminelli:2008es} for vacuum fluctuations about a cosmological solution.  This is not surprising since we are looking at different effects. Tunneling is itself a non-perturbative phenomena, and so the perturbative bounds derived there are not appropriate. Tunneling is described by rare stochastic fluctuations, and the conditions for the validity of the stochastic tunneling calculation are necessarily more restrictive for these rare fluctuations, in comparison to the typical fluctuations about cosmological solutions considered in Refs.~\cite{Leblond:2008gg, Creminelli:2008es}.

Returning to the tunneling rate, which as we have argued is principally determined by the irreversibility ratio ${\cal R}$ we find that the 
correction to $\ln \Gamma$ from the non-Gaussian term takes the form
\be
\delta \ln \Gamma \sim \langle f_{NL}\rangle  \ln \Gamma,
\ee
with similar contributions from higher $N$-point functions. To get a reliable result, we would have to sum 
this full set of contributions. For minimally coupled scalars we always have $\langle f_{NL} \rangle \sim \langle \epsilon_{\rm slow\, roll}\rangle \ll 1$ and so these higher order corrections are negligible. This explains why in the standard case there is perfect agreement between the local and global descriptions of the tunneling process \cite{Lindebook}.

\subsection{A contradiction?}

The fact that the effect of the non-gaussian noise correlation functions become significant during
tunneling is evidence that the {\it dynamics} of tunneling in DBI inflation
is strongly coupled, i.e. strongly sensitive to the higher order kinetic terms. Since the strongly coupled
regime is difficult or impossible to treat perturbatively, we should not be 
surprised that a perturbative expansion of the logarithm of the influence functional is insufficient to capture these dynamics fully. In practice this means that the familiar picture of stochastic inflation with gaussian noise is no longer appropriate for DBI inflation, at least in the description of tunneling.

However, the Euclidean calculation gives the tunneling rate independent of
the tunneling trajectory's dynamics to be
\be
\Gamma  \sim \exp\( {S_{\rm ent}(\phi_{\rm barrier})-S_{\rm ent}(\phi_{\rm false})}\), \quad \quad S_{\rm ent} = \frac{8 \pi^2 M_{\rm pl}^2}{H^2(\phi)}.
\ee
This result is identical to that for inflation driven by a minimally coupled scalar field and seems to be insensitive to the nontrivial kinetic terms in the DBI action, at least as long as we satisfy $H^4 \ll T(\phi_{\rm barrier})$. If these two calculations are really describing the same physics, it is difficult to see how it could be the case that one rate is essentially insensitive to the non-minimal kinetic terms, whilst the other calculation is strongly sensitive to the new interactions that arise in the presence of these terms.

The probable resolution is that the while the full tunneling dynamics is strongly coupled, the Euclidean calculation gets the correct answer because it correctly captures the underlying gravitational thermodynamics of the problem. Thermodynamics tells us that the rate is independent of the path chosen, and thus is independent of the fact that $c_s$ may depart from unity along the tunneling path. Thus the Euclidean calculation innocently gets the correct answer without really capturing the full non-equilibrium dynamics. Since gravitational thermodynamics is likely to be fundamental, it seems reasonable to suppose that if the contribution from all the higher order noise correlation functions could somehow be resummed, the result for the irreversibility ratio ${\cal R}$ would ultimately be consistent with the fluctuation theorem, and hence agree with the standard Hawking-Moss result, as long as the barrier remains in the regime where the fluctuation determinant can be well approximated by a one-loop gaussian integral.

\subsection{Consequences for other instanton tunneling rates}

Another recent calculation of the consequences for tunneling -- in the Coleman-de Luccia
instanton, described in Ref. \cite{Brown:2007ce} -- found a sound speed
enhancement to that quantum tunneling rate. 
As emphasized by \cite{Brown:2007sd, Hackworth:2004xb}, though, the Hawking-Moss instanton is the extreme case in which tunneling takes place via a maximally broad ``bounce", or in a very thick tunneling ``bubble".  The instanton calculation in this case is almost entirely determined
by the evaluation of the action exactly at the unstable maximum on the tunneling potential,
where the non-trivial sound speed effects are suppressed. In the opposite extreme,
Coleman-de Luccia, the quantum tunneling is rapid and proceeds for typically subhorizon sized bubbles. Ref. \cite{Hackworth:2004xb}
demonstrates that there are typically a large number of interpolating instantons between Hawking-Moss and Coleman-de Luccia in the case of canonical scalar fields. It is natural to expect
that each of these interpolating instantons will also exist for DBI and other non-minimal kinetic theories. Though we have not yet pursued the question, we expect DBI effects will begin to
show up in these other instantons, but will become progressively smaller as the bubbles get broader and we approach the Hawking-Moss instanton.

\section{Summary}

The thermal tunneling rate described by the Hawking-Moss instanton is apparently insensitive to the strong coupled dynamics of theories with non-minimal kinetic terms as long as the barrier lies in the regime where the fluctuation determinant can be well approximated by a one-loop gaussian integral. However, the actual non-equilibrium dynamics that a tunneling 
trajectory follows is strongly coupled whenever $c_s$ departs sufficiently from unity along the tunneling trajectory. The calculable estimate of the stochastic tunneling rate, assuming gaussian noise from subhorizon fluctuations, disagrees with the Hawking-Moss
calculation in the $\langle c_s \rangle \ll 1$ case. We have argued that the stochastic tunneling rate and the thermal Hawking-Moss tunneling rate should be complementary descriptions of the same
physics, and consequently it is a surprise that they should appear to disagree.
The probable resolution of this apparent conflict is that the strongly coupled physics
must, when resummed, respect the underlying gravitational thermodynamics
that govern de Sitter space and which are manifest in the local picture. We have explicitly demonstrated that the non-gaussian noise fluctuations induced by the subhorizon modes become order unity effects during the tunneling process, negating a perturbative calculation of the stochastic tunneling rate.
Since the fluctuation theorem, the fundamental statement of thermodynamics,
agrees with the Hawking-Moss instanton, we expect this to be the ultimate answer,
which implies that the full rate of tunneling is blind to the strongly coupled 
dynamics of the tunneling path. We leave it to future work to hopefully provide a fully non-equilibrium derivation of this result and demonstrate the consistency between the local and global descriptions of tunneling.

\acknowledgments
We thank Sarah Shandera and the anonymous referee for questions and comments that
greatly enhanced the clarity of the text and L. Susskind, B. Freivogel, A. Brown, H. Tye,
and S. Sarangi for helpful discussions.
The work of A.J.T. and M.W.  was supported by the Perimeter Institute for Theoretical
Physics.  Research at the Perimeter Institute is supported by the Government
of Canada through Industry Canada and by the Province of Ontario through
the Ministry of Research \& Innovation.


\begin{thebibliography}{19}

\bibitem{DBI} 
  E.~Silverstein and D.~Tong,
  ``Scalar speed limits and cosmology: Acceleration from D-cceleration,''
  Phys.\ Rev.\  D {\bf 70}, 103505 (2004)
  [arXiv:hep-th/0310221];\\
  M.~Alishahiha, E.~Silverstein and D.~Tong,
  ``DBI in the sky,''
  Phys.\ Rev.\  D {\bf 70}, 123505 (2004)
  [arXiv:hep-th/0404084].

\bibitem{kinflation} 
  C.~Armendariz-Picon, T.~Damour and V.~F.~Mukhanov,
  ``k-inflation,''
  Phys.\ Lett.\  B {\bf 458}, 209 (1999)
  [arXiv:hep-th/9904075];\\
  J.~Garriga and V.~F.~Mukhanov,
  ``Perturbations in k-inflation,''
  Phys.\ Lett.\  B {\bf 458}, 219 (1999)
  [arXiv:hep-th/9904176].

\bibitem{kessence} 
  T.~Chiba, T.~Okabe and M.~Yamaguchi,
  ``Kinetically driven quintessence,''
  Phys.\ Rev.\  D {\bf 62}, 023511 (2000)
  [arXiv:astro-ph/9912463];\\
    C.~Armendariz-Picon, V.~F.~Mukhanov and P.~J.~Steinhardt,
  ``A dynamical solution to the problem of a small cosmological constant  and
  late-time cosmic acceleration,''
  Phys.\ Rev.\ Lett.\  {\bf 85}, 4438 (2000)
  [arXiv:astro-ph/0004134];\\
  C.~Armendariz-Picon, V.~F.~Mukhanov and P.~J.~Steinhardt,
  ``Essentials of k-essence,''
  Phys.\ Rev.\  D {\bf 63}, 103510 (2001)
  [arXiv:astro-ph/0006373].
  
\bibitem{Creminelli1} 
  C.~Cheung, P.~Creminelli, A.~L.~Fitzpatrick, J.~Kaplan and L.~Senatore,
  ``The Effective Field Theory of Inflation,''
  JHEP {\bf 0803}, 014 (2008)
  [arXiv:0709.0293 [hep-th]].


\bibitem{Brown:2007ce}
  A.~R.~Brown, S.~Sarangi, B.~Shlaer and A.~Weltman,
  ``A Wrinkle in Coleman - De Luccia,''
  Phys.\ Rev.\ Lett.\  {\bf 99}, 161601 (2007)
  [arXiv:0706.0485 [hep-th]].

\bibitem{Tolley:2008na}
  A.~J.~Tolley and M.~Wyman,
  ``Stochastic Inflation Revisited: Non-Slow Roll Statistics and DBI
  Inflation,''
  arXiv:0801.1854 [hep-th].

\bibitem{related}

  X.~Chen, S.~Sarangi, S.~H.~Henry Tye and J.~Xu,
  ``Is brane inflation eternal?,''
  JCAP {\bf 0611}, 015 (2006)
  [arXiv:hep-th/0608082].

\bibitem{related2}

  F.~Helmer and S.~Winitzki,
  ``Self-reproduction in k-inflation,''
  Phys.\ Rev.\  D {\bf 74}, 063528 (2006)
  [arXiv:gr-qc/0608019].

\bibitem{Gibbons:1977mu}
  G.~W.~Gibbons and S.~W.~Hawking,
  ``Cosmological Event Horizons, Thermodynamics, And Particle Creation,''
  Phys.\ Rev.\  D {\bf 15}, 2738 (1977).

\bibitem{Susskind}
 
   L.~Dyson, M.~Kleban and L.~Susskind,
  ``Disturbing implications of a cosmological constant,''
  JHEP {\bf 0210}, 011 (2002)
  [arXiv:hep-th/0208013];\\


\bibitem{Brown:2007sd}
  A.~R.~Brown and E.~J.~Weinberg,
  ``Thermal derivation of the Coleman-De Luccia tunneling prescription,''
  Phys.\ Rev.\  D {\bf 76}, 064003 (2007)
  [arXiv:0706.1573 [hep-th]].

\bibitem{emergent} 
E.~Babichev, V.~Mukhanov and A.~Vikman,
  ``k-Essence, superluminal propagation, causality and emergent geometry,''
  JHEP {\bf 0802}, 101 (2008)
  [arXiv:0708.0561 [hep-th]].

\bibitem{Leblond:2008gg}
  L.~Leblond and S.~Shandera,
  ``Simple Bounds from the Perturbative Regime of Inflation,''
  arXiv:0802.2290 [hep-th].
  
\bibitem{Creminelli:2008es}

  N.~Arkani-Hamed, S.~Dubovsky, A.~Nicolis, E.~Trincherini and G.~Villadoro,
  ``A Measure of de Sitter Entropy and Eternal Inflation,''
  JHEP {\bf 0705}, 055 (2007)
  [arXiv:0704.1814 [hep-th]].


  P.~Creminelli, S.~Dubovsky, A.~Nicolis, L.~Senatore and M.~Zaldarriaga,
  ``The Phase Transition to Slow-roll Eternal Inflation,''
  arXiv:0802.1067 [hep-th].
  
  
\bibitem{Batra:2006rz}
  P.~Batra and M.~Kleban,
  ``Transitions Between de Sitter Minima,''
  Phys.\ Rev.\  D {\bf 76}, 103510 (2007)
  [arXiv:hep-th/0612083].


\bibitem{FT} D.~J.~Evans and D.~J.~Searles, ``The Fluctuation Theorem", Advances in Physics  51 (2002) 1529Ð1585.

  
\bibitem{Wohns:2008qv}
  D.~Wohns,
  ``Hawking-Moss Tunneling with a Dirac-Born-Infeld Action,''
  arXiv:0802.0623 [hep-th].

\bibitem{Lindebook}

  A.~D.~Linde,
  ``Particle Physics and Inflationary Cosmology,'' Harwood, Switzerland (1990),
  arXiv:hep-th/0503203.


\bibitem{Hackworth:2004xb}
  J.~C.~Hackworth and E.~J.~Weinberg,
  ``Oscillating bounce solutions and vacuum tunneling in de Sitter
  spacetime,''
  Phys.\ Rev.\  D {\bf 71}, 044014 (2005)
  [arXiv:hep-th/0410142].

\end{thebibliography}
\end{document}